\documentstyle[amssymb,aps,preprint,12pt]{revtex}
\begin{document}
\title{Effect of surface tension and depolarization field on ferroelectric
nanomaterials properties}
\author{M.D.Glinchuk, A.N.Morozovskaya}
\address{Institute for Problems of Materials Science, NASc of Ukraine, \\
Krjijanovskogo 3, 03680 Kiev-142, Ukraine}
\maketitle

\begin{abstract}
The theory of size effects of the properties of nanocrystalline
ferroelectric ceramic or nanoparticle powder allowing for surface tension
and depolarization field is proposed. Surface tension was included into free
energy functional and surface energy was expressed via surface tension
coefficient. The latter was shown to be dependent on temperature due to its
relation to dielectric permittivity of the nanoparticles. The depolarization
field effect was calculated in the model taking into account the space
charge layer on the surface, this space-charge being able to compensate
depolarization field in the bulk material.

Euler-Lagrange Equation for inhomogeneous polarization of nanomaterial with
boundary condition where extrapolation length was shown to be temperature
dependent quantity was solved analytically both in paraelectric and
ferroelectric phase of size driven phase transition. This phase transition
critical temperature dependence on the particle size was calculated.
Temperature and size dependence of nanomaterials polarization and dielectric
susceptibility was obtained. The possibility to calculate these and other
properties by minimization of conventional free energy in the form of
different power polarization series, but with the coefficients which depend
on particles size, temperature, contribution of depolarization field and
surface tension coefficient was demonstrated. These latter effects were
shown to influence essentially the nanomaterial properties. The comparison
with available experimental data is performed.
\end{abstract}

\section{Introduction}

The characteristic feature of ferroelectric nanomaterials (e.g. the ceramics
with nanosize grains or powder built by nanosize particles) is known to be
inhomogeneous polarization, dielectric susceptibility and other properties
(see e.g. ref. \cite{WLZ1}, \cite{Niep}, \cite{Rag} and ref. therein). The
properties inhomogeneity is related to the strong influence of the surface,
where the properties differ from those in the bulk. Because of this surface
and correlation energy (polarization gradient $\nabla P$) have to be
included when calculating the nanomaterials properties. In phenomenological
approach for ferroelectrics with perofskite structure the additional terms
to free energy density have the form $\delta (\nabla P)^2/2$ and $\delta
P_s^2/2\lambda $, where $P_s$ is polarization on the surface, $\lambda $ and 
$\delta $ are material constants. The variation of free energy functional is
usually performed to obtain equilibrium polarization. This leads to
Euler-Lagrange equation for the polarization with boundary conditions, which
include the extrapolation length $\lambda $ (see e.g. ref. \cite{WLZ2}) as
combined characteristic of surface energy and polarization gradient.
Moreover, the extrapolation length value reflects the degree of the
polarization inhomogeneity. It is because the boundary condition has the
form of tangent line to the polarization on the surface (e.g. for spherical
particle with radius $R$ the boundary condition is $\left( \lambda \frac{dP}{%
dr}+P\right) _{r=R}=0$), so that the smaller homogeneity of polarization the
larger the extrapolation length and vise versa. Therefore it is important to
calculate $\lambda $ value rather than to consider it as phenomenological
parameter. While parameter $\delta $ can be estimated via the material
correlation radius nothing is known about $\lambda $ value. On the other
hand calculation of $\lambda $ in Ising model includes mainly the change of
the average surface coordinate number (see e.g. ref.\cite{WLZ2}, \cite{WaSm}
and ref. therein) that seems to be oversimplification for $\lambda $ value
estimation.

Another problem is related to contribution of depolarization field effect.
The influence of this field that is able to destroy spontaneous polarization
even in bulk materials can be decreased by such internal factors as free
carriers and domain structure and it can be completely compensated under
short circuit conditions. It was shown that inhomogeneous polarization in
the ferroelectric films makes impossible to compensate the depolarization
field in short circuited films made from single domain ferroelectric, that
is perfect insulator \cite{6}. However in the majority of works the
influence of depolarization field in nanomaterials has not been considered
yet. In ref. \cite{7} the influence of depolarization field was considered
in the model of the multidomain ferroelectric of finite size and
space-charge layers on the surfaces. However the surface energy of the
particles was not taken into account, which is known to play the decisive
role in the physics of nanomaterials. The important role of the surface
tension was demonstrated in \cite{Wenhui}, where observed size dependence of
the phonon modes frequencies in nanocrystalline PbTiO$_{3}$ was shown to be
attributed to the surface tension.

In the present work we propose the model for consideration of depolarization
field effect. We present the surface energy via surface tension that made it
possible to express the extrapolation length via surface tension
coefficient. This coefficient has been shown to be dependent on temperature $%
T$, that leads to extrapolation length dependence on $T$ and to appearance
of some additional temperature dependence of the properties. The
ferroelectric nanoparticles are supposed to be single domain (e.g.
polarization $P\equiv P_{z}$) with space-charge layer on the surface. The
solution of Euler-Lagrange equation for polarization both in paraelectric
and ferroelectric phases has shown the possibility to obtain the
nanomaterials properties from free energy expression in the form of power
series of polarization like that in the bulk, but with renormalized
coefficients. These coefficients depend on temperature and the particles
sizes. Although there is a distribution of these sizes in nanomaterials, so
that the properties have to be averages with the sizes distribution
function, we consider only the average radius $R$ of the particles. In many
experimental works (see e.g. ref.\cite{WLZ1}, \cite{Niep}) the properties of
nanomaterials were measured as a function of a particle average size. The
average size can be determined from integrated width of diffraction peak
using Scherrer's formula or with the help of TEM method. Both results have
been shown to be in a fairly good agreement with each other \cite{WLZ1}.

\section{Surface energy and surface tension}

\subsection{Surface energy}

Surface energy is known to be the energy of surface tension on the boundary
between two media (see e.g. ref.\cite{Stat}). It was included into
thermodynamic potential written for ferroelectric nanomaterial in ref. \cite
{9} both as a surface energy and as hydrostatic pressure, acting on
nanoparticles. However these terms related to surface tension were not
expressed via electric polarization (order parameter of ferroelectrics), so
that it appeared impossible to incorporate them directly into Landau free
energy that has the form of power series of polarization. This incorporation
seems to be important to pour light on physical meaning of some parameters
of thermodynamic theory approach to nanomaterial properties calculation.

In what follows we will amend this limitation of the previous consideration
of surface tension contribution to the physical properties of nanomaterials.
In general case the surface energy can be written in the form \cite{Stat}:

\begin{equation}
E_{surf}=\int \mu ds  \eqnum{1}
\end{equation}
where $\mu $ is the surface tension coefficient and integration is performed
over the surface. The small change of the surface area $ds$ can be rewritten
by the following way

\begin{equation}
ds=U_{xx}U_{yy}dxdy  \eqnum{2}
\end{equation}

Here $U_{ii}$ ($i=x,y$) is a component of strain tensor, the normal to the
surface is oriented along $z$ axis. Keeping in mind that there is no
inversion center near the material surface, the strain tensor component can
be related to surface polarization $P_{sz}$ via piezoelectric coefficients $%
d_{ii}$, namely $U_{ii}=d_{iiz}P_z$. Substitution of this relation into Eqs.
(1), (2) yields:

\begin{equation}
E_{surf}=\int \mu d_{xxz}d_{yyz}P_{sz}^2dxdy  \eqnum{3}
\end{equation}

Note that Eq. (3) can be used directly for ferroelectric thin films also.
Contrary to the films in nanomaterials there is another contribution from
surface tension related to the uniform mechanical pressure $p=\mu
/(R_{1}+R_{2})$, where $R_{1}$ and $R_{2}$ are the main curvature radiuses
of the particle surface ($R_{1,2}\rightarrow \infty $ for the films). For
ferroelectrics without inversion center this pressure induces polarization
via piezoelectric coupling. For ferroelectrics with inversion center (e.g.
ferroelectrics with perofskite structure) the contribution of the pressure
can be expressed via electrostriction and included into free energy on the
base of the procedure proposed in \cite{PeTa}. Therefore free energy (with
determined stress $\sigma _{ik}=-\delta _{ik}p$, expressed in terms of
uniform pressure $p=\mu /(R_{1}+R_{2})$, polarization $P_{z}$ and strain $%
U_{ik}=\delta _{ik}U$ as independent variables) acquires the form: 
\begin{eqnarray}
F &=&\int d^{3}r\left[ \frac{\alpha (T-T_{c})}{2}P_{z}^{2}+\frac{\beta }{4}%
P_{z}^{4}+\frac{\gamma }{6}P_{z}^{6}+\frac{\delta }{2}\left( \nabla
P_{z}\right) ^{2}-\frac{3}{2}C_{11}U_{h}^{2}\right. -  \eqnum{4a} \\
&&-\left.
3C_{12}\,U_{h}^{2}-Q_{11}U_{h}P_{z}^{2}-2Q_{12}U_{h}P_{z}^{2}-E_{z}P_{z}-\mu
\left( \frac{1}{R_{1}}+\frac{1}{R_{2}}\right) U_{h}\right] +E_{surf} 
\nonumber
\end{eqnarray}
where $T_{c}$ is ferroelectric transition temperature in a bulk material, $%
Q_{ij}$ and $C_{ij}$ are respectively electrostriction and elastic
coefficients in Voigt presentation, $E_{z}$ is the electric field. For
spherical particles $R_{1}=R_{2}=R$. When taking into account the particle
spherical symmetry, we assume that the magnitude of oriented along $z$ axis
polarization $P(r)\equiv P$ depends only upon the radial position $r$.
Analogically to \cite{PeTa} and taking into account the depolarization field
energy, one can obtain the renormalized free energy in the form

\begin{eqnarray}
F &=&4\pi \int r^{2}dr\left[ \frac{\widetilde{a}}{2}P^{2}+\frac{b}{4}P^{4}+%
\frac{c}{6}P^{6}+\frac{\delta }{2}\left( \frac{dP}{dr}\right) ^{2}-\left( 
\frac{1}{2}E_{d}+E_{0}\right) P\right] +8\pi \int \mu d^{2}P^{2}rdr 
\eqnum{4b} \\
\widetilde{a} &=&\alpha (T-T_{c})+\frac{4\mu }{R}\frac{Q_{11}+2Q_{12}}{%
C_{11}+2C_{12}},\qquad b=\beta +\frac{Q_{11}+2Q_{12}}{6(C_{11}+2C_{12})}%
,\qquad c=\gamma ,  \nonumber
\end{eqnarray}
Here $E_{0}$ is external electric field, $E_{d}$ is depolarization field $%
d_{xxz}=d_{yyz}\equiv d$ is inverse piezoelectric effect constant.
Hereinafter we will consider ferroelectrics with perofskite structure and
nanoparticles of spherical form so we will use the Eq. (4b).

\subsection{Surface tension}

Let us proceed to the consideration of surface tension coefficient
temperature dependence related to phase transition from paraelectric to
ferroelectric phase. Critical temperature dependence of surface tension
coefficient when approaching the transition from solid state to liquid state
is well known \cite{Stat}. To the best of our knowledge nobody considered
influence of structural phase transition like paraelectric-ferroelectric on
surface tension. We carried out the calculations in the following model.

In the solids the appearance of polarization manifests itself via the
charges on the surface. In nanomaterials and in the films the charges can
appear in some region near the surface where polarization is essentially
inhomogeneous. It is known that these charges cause the depolarization field
in ferroelectrics. In the considered case of single domain nanoparticles the
only possibility of depolarization field compensation is the space-charge
layer on the surface that has to include the charges of opposite sign to
those related to polarization. The possible sources of space-charge layer
including free carriers (if any) localized near the surface one can find in
ref. \cite{7}. The influence of the flat surface on the aforementioned
charges $q$ can be considered by means of the interaction with image charges 
$q^{*}=q(\varepsilon _i-\varepsilon _e)/\left( \varepsilon _i+\varepsilon
_e\right) $ (see, e.g. ref.\cite{Elect}), where $\varepsilon _i$ and $%
\varepsilon _e$ are respectively the dielectric permittivity of the material
of particle and surrounding media. All these charges cause the additional
energy $w_s$ related to the surface \cite{Stat}, so that

\begin{equation}
\Delta \mu _s=n\int\limits_Vw_s(x,y,z)d^3r  \eqnum{5a}
\end{equation}
Here $n$ is the concentration of the charges related to inhomogeneous
spontaneous polarization (i.e. it is non-zero only in ferroelectric phase),
the integration is performed over the volume of particle. Taking into
account Coulomb interaction between the dipole with moment $r_d\,q$ and its
image, one obtains the following expression

\begin{equation}
\Delta \mu _s=\frac{n\,q\,q^{*}}{\varepsilon _i}\int\limits_a^{l\gg a}\left( 
\frac 1{2\,z}+\frac 1{2\,z+2\,r_d}-\frac 2{2\,z+r_d}\right) dz  \eqnum{5b}
\end{equation}
Here $a$, $r_d$ and $l$ are respectively a lattice constant, minimal
distance between positive and negative charges and the size of the region of
inhomogeneous polarization. Integration in Eq. (5) yields

\begin{equation}
\Delta \mu _s=\frac{n\ q^2}2\frac{\varepsilon _i-\varepsilon _e}{\varepsilon
_i(\varepsilon _i+\varepsilon _e)}\ln \left( \frac{4a(a+r_d)}{(2a+r_d)^2}%
\right)  \eqnum{6a}
\end{equation}
This relation can be simplified in supposition $r_d/a<<1$, so that

\begin{equation}
\Delta \mu _s\approx \frac{n\ q^2}8\frac{\varepsilon _i-\varepsilon _e}{%
\varepsilon _i(\varepsilon _i+\varepsilon _e)}\left( \frac{r_d}a\right) ^2 
\eqnum{6b}
\end{equation}

It should be noted that the consideration we just have carried out can be
applied both for flat surface and for the surfaces with the principal radii
of curvature much larger than $r_d$ value. This statement follows from the
derivation of $\Delta \mu _s$ for the case if spherical particle performed
in Appendix1.

It should be underlined that since for spherical particles $\Delta \mu
_{s}(T)\sim (\varepsilon _{i}-\varepsilon _{e})/\left( \varepsilon
_{i}(\varepsilon _{i}+\varepsilon _{e})\right) $ the temperature dependence
(see below) is the same for all the considered forms. Moreover since
approximately $r_{d}/a\leq 1$, it is possible to say that the expressions
(6a) and (A1-6) coincide with each other. More generally it would be better
to rewrite{\em \ }$\Delta \mu _{s}$\ in the following form{\em :} 
\begin{equation}
\Delta \mu _{s}=\varkappa \ n\ q^{2}\frac{\varepsilon _{i}-\varepsilon _{e}}{%
\varepsilon _{i}(\varepsilon _{i}+\varepsilon _{e})}  \eqnum{6c}
\end{equation}
where the value of the coefficient $\varkappa $ depends on particle shape,
its value being $\varkappa \leq 1$.

The surface tension coefficient in ferroelectric phase can represented as
the sum of $\Delta \mu _s$ and that of paraelectric phase $\mu _0$, i.e.

\begin{equation}
\mu =\mu _0+\Delta \mu _s,  \eqnum{7}
\end{equation}

One can see, that while $\mu _{0}$ is temperature independent, $\Delta \mu
_{s}$ depends on temperature mainly via $\varepsilon _{i}(T)$. Really in the
considered cases of nanomaterials in the form of nanopowders or
nanocrystalline ceramics the media surrounding nanoparticles may be
considered as non-ferroelectric one. In such a case at phase transition
temperature $\varepsilon _{i}(T_{cr})\rightarrow \infty $ and so $\Delta \mu
_{s}\rightarrow 0$, $\mu \rightarrow \mu _{0}$ (see (6), (7)). Note, that in
the case of two ferroelectric media (e.g. BaTiO$_{3}$ particles dispersed in
SrTiO$_{3}$ or PbZr$_{1-x}$Ti$_{x}$O$_{3}$ media) both $\varepsilon _{i}$
and $\varepsilon _{e}$ depend on $T$. If the transition temperature in
surrounding media $T=T_{ec},$ in these composite materials $\Delta \mu
_{s}\rightarrow -n\ q^{2}/\varepsilon _{i}(r_{d}/a)^{2}$ at $T\rightarrow
T_{ec}$. Note, that the negative sign of $\Delta \mu _{s}$ can be valid in
some temperature region where $\varepsilon _{e}>\varepsilon _{i}$ in both
aforementioned cases. Keeping in mind that surface tension must be positive,
it is necessary to add the inequality $\left| \Delta \mu _{s}\right| /\mu
_{0}<1$ to Eq.(7).

Hereinafter we consider only the case of nonferroelectric surrounding media.

\section{Depolarization field effect}

We will consider this effect in the model described in previous section.
Namely the charges localized in the thin surface layer, where polarization
is the most inhomogeneous, create depolarization field inside the particle.
This field can be compensated by space-charge layer on the surface (see
Section 2).

The magnitude of depolarization field was calculated on the base of equation
written in ref. \cite{Elect} for homogeneously polarized dielectric
submerged in a medium with dielectric susceptibility $\varepsilon _e$. To
take into account additional charges related to inhomogeneous polarization
and those from space-charge layer we added the term $-4\pi n_\alpha
P_{V\alpha }$. In the case of external field absence this yields

\begin{equation}
\left[ (1-n_\alpha )\varepsilon _e+n_\alpha \right] E_{d\alpha }+4\pi
n_\alpha (P_\alpha -P_{V\alpha })=0,~\alpha =x,y,z  \eqnum{8}
\end{equation}
where $E_d$ is depolarization field, $n_\alpha $ is depolarization factor,
that depends on the sample form and for spherical form $n_\alpha =1/3$ (see
e.g. ref.\cite{Elect}), all the quantities $E_{d\alpha }$, $P_\alpha $, $%
P_{V\alpha }$ being dependent on the coordinates of considered point $%
\overrightarrow{r}$. The $P_{V\alpha }$ will be considered as variational
parameter to minimize the energy related to depolarization field.

With respect to Eq. (8) this energy can be written as

\begin{equation}
G=\frac 1V\frac 12\sum\limits_\alpha \int E_{d\alpha }^2dV=\frac{8\pi ^2}V%
\sum\limits_\alpha \left( \frac{n_\alpha }{(1-n_\alpha )\varepsilon
_e+n_\alpha }\right) ^2\int (P_\alpha -P_{V\alpha })^2dV  \eqnum{9}
\end{equation}

Equilibrium value of $P_{V\alpha 0}$ can be obtained from the condition $%
\frac{\delta G}{\delta P_{V\alpha }}=0$, that yields

\begin{equation}
P_{V\alpha 0}=\frac 1V\int dVP_\alpha (\overrightarrow{r})\equiv \overline{P}%
_\alpha  \eqnum{10}
\end{equation}
where $\overline{P}$ is average polarization.

The Eq. (8) with respect to Eq. (10) gives the following equation for
depolarization field:

\begin{equation}
E_{d\alpha }=-\frac{4\pi n_\alpha (P_\alpha -\overline{P}_\alpha )}{%
(1-n_\alpha )\varepsilon _e+n_\alpha }  \eqnum{11}
\end{equation}

For single domain spherical nanoparticle with polarization along $z$ axis
(i.e. $\alpha =z$, $n_z=1/3$) Eq. (11) gives

\begin{equation}
E_{dz}=-a_0(P_z-\overline{P}_z),~a_0\equiv \frac{4\pi }{1+2\varepsilon _e} 
\eqnum{12}
\end{equation}

Note, that for the film where $z$ coincides with normal to the film surface $%
n_z=1$ \cite{Elect}, so that Eq. (11) leads to the expression obtained in
ref.\cite{6} for short circuited film: $E_{dz}=-4\pi (P_z-\overline{P}_z)$.
The later is related to the fact that both the short circuit condition and
the charges on the surface produce similar potential distribution which is
able to compensate the depolarization field completely (when polarization is
homogeneous, i.e. $P_z=\overline{P}$) or partly (when polarization is
inhomogeneous). Note, that Eq. (11) makes it possible to calculate the
depolarization field for nanoparticles of different form for which
depolarization factors $n_\alpha $ are known. In particular for the particle
of cylindrical form with $z$ along the axis of rotation $n_z=0$, i.e. $%
E_{dz}=0$.

\section{Free energy functional and Euler-Lagrange equation for polarization}

\subsection{Free energy}

Allowing for the results of previous sections for depolarization field
contribution (see Eq. (12)), we can write the free energy (4b) in the
following form:

\begin{equation}
F=\frac 1V\left( 4\pi \int\limits_0^Rr^2dr\left[ \frac a2P^2+\frac b4P^4+%
\frac c6P^6+\frac \delta 2\left( \frac{dP}{dr}\right) ^2-\left( \frac{a_0}2%
\overline{P}+E_0\right) P\right] +8\pi \int\limits_0^R\mu d^2P^2rdr\right) 
\eqnum{13}
\end{equation}

\begin{equation}
a=\alpha (T-T_c)+\frac{4\mu }R\frac{Q_{11}+2Q_{12}}{C_{11}+2C_{12}}+a_0 
\eqnum{14}
\end{equation}

Having written Eq. (13) we take into account spherical symmetry and that is
why the magnitude of oriented along $z$ axis polarization $P(r)\equiv P$
depends only upon the radial position $r$. By virtue of surface tension
temperature dependence (see Eqs. (6), (7)) additional terms with temperature
dependence appear in surface energy (see last term in Eq. (13)) and in the
coefficient before $P^{2}$ (see Eq. (14)). Since $\mu >0$ and $%
Q_{11}+2Q_{12}>0$ in perofskites both the second term in $a$ (related to
surface tension) and the third one (related to depolarization field) tends
to decrease $T_{c}$ value, i.e. to destroy ferroelectric phase transition.

\subsection{Euler-Lagrange equation}

Variation of free energy functional (13) yields the following Euler-Lagrange
equation for static polarization distribution and the boundary condition:

\begin{eqnarray}
aP+bP^3+cP^5-\delta \left( \frac{d^2P}{dr^2}+\frac 2r\frac{dP}{dr}\right)
&=&E_0+a_0\overline{P}  \eqnum{15a} \\
\left( \lambda \frac{dP}{dr}+P\right) _{r=R} &=&0  \eqnum{15b} \\
\lambda &=&\frac \delta {\mu d^2}  \eqnum{15c}
\end{eqnarray}

It follows from Eq. (15c), that extrapolation length $\lambda $ depends on
surface tension coefficient $\mu (T)$ given by Eqs. (6), (7), so that $%
\lambda $ has to be temperature dependent quantity. On the other hand $\mu
>0 $ in all the materials \cite{Stat}, $d^2>0$ so the sign of $\lambda $
defines by that of $\delta $. In what follows we will consider the case $%
\delta >0$ that leads to positive extrapolation length $\lambda >0$. Note
that in the case $\delta <0$ one has to add the fourth power of gradient
into free energy expansion to conserve the system stability. It is seen that
it is possible to control extrapolation length magnitude (i.e. boundary
condition)by the choice of the materials with the smaller or larger
piezoelectric constant and surface tension.

\section{Size driven ferroelectric phase transition}

Let us begin with the solution of Eq. (15a) with boundary condition (15b) in
paraelectric phase where the polarization induced by external electric field
can be small enough, so that the nonlinear terms in Eq. (15a) can be
neglected. In such a case the substitution of $P=\widetilde{P}/r$ transforms
Eqs. (15a), (15b) to the following forms:

\begin{eqnarray}
a\frac{\widetilde{P}}r-\delta \frac 1r\frac{d^2\widetilde{P}}{dr^2}
&=&E_0+a_0\overline{P}  \eqnum{16a}  \label{16a} \\
\lambda \left. \frac{d\widetilde{P}}{dr}\right| _{r=R}+\left( 1-\frac \lambda
R\right) \left. \widetilde{P}\right| _{r=R} &=&0  \eqnum{16b}  \label{16b}
\end{eqnarray}

Keeping in mind, that in the particle center polarization $P(0)$ is finite,
one can easily obtain the solution of Eq. (16a) and write $P(r)$ as

\begin{eqnarray}
P(r) &=&\frac 1a(E_0+a_0\overline{P})\left[ 1-\frac Rr\frac{sh(r\sqrt{%
a/\delta })}{M(R)}\right]  \eqnum{17a}  \label{17a} \\
M(R) &=&\lambda \sqrt{\frac a\delta }ch\left( R\sqrt{\frac a\delta }\right)
+\left( 1-\frac \lambda R\right) sh\left( R\sqrt{\frac a\delta }\right) 
\eqnum{17b}  \label{17b}
\end{eqnarray}

Averaged polarization can be obtained by the integration of (17a) over
particle volume, i.e. $\overline{P}=3/R^3\int\limits_0^RP(r)r^2dr$, that
yields

\begin{eqnarray}
\overline{P}(R) &=&\frac{E_0W(R)}{a-a_0W(R)}  \eqnum{18a}  \label{18a} \\
W(R) &=&1-\frac 3{R^2}\frac \delta a\frac{R\sqrt{a/\delta }ch(R\sqrt{%
a/\delta })-sh(R\sqrt{a/\delta })}{M(R)}  \eqnum{18b}  \label{18b}
\end{eqnarray}

Substituting Eq. (18a) into (17a), one obtains

\begin{equation}
P(r)=\frac{E_0}{a-a_0W(R)}\left[ 1-\frac Rr\frac{sh(r\sqrt{a/\delta })}{M(R)}%
\right]  \eqnum{19a}  \label{19a}
\end{equation}

It is clearly seen that polarization in paraelectric phase is proportional
to external field, the coefficient of proportionality being the
inhomogeneous dielectric susceptibility

\begin{equation}
\chi _{PE}(r)=\frac 1{a-a_0W(R)}\left[ 1-\frac Rr\frac{sh(r\sqrt{a/\delta })%
}{M(R)}\right]   \eqnum{19b}  \label{19b}
\end{equation}

It is follows from Eqs. (18a,b) and (14) that for bulk material ($%
R\rightarrow \infty $) $W(R)\rightarrow 1$ and $\overline{P}(R)\rightarrow
E_0/(\alpha (T-T_c))$ as it has to be.

The average value of susceptibility can be extracted from Eq. (18a) as the
coefficient before $E_0$. It follows from Eq. (18a) that the susceptibility
becomes infinitely large at the condition:

\begin{equation}
a-a_0W(R)=0  \eqnum{19c}
\end{equation}

Allowing for that both $a$ and $W(R)$ depend on particle radius $R$ and
temperature $T$, the condition (19c)\ can be satisfied at some critical
radius $R_{cr}$ (temperature is fixed) or at some critical temperature $%
T_{cr}$ (radius is fixed). These critical values correspond to size driven
ferroelectric phase transition in nanomaterials. Its consideration without
contribution of depolarization field and surface tension was carried out
earlier in e.g. ref. \cite{WLZ2}, \cite{WaSm}, \cite{YGW}, \cite{Rych}.
Comparative analysis of the results with and without these factors
contribution will be performed later.

Because of complex enough form of $W(R)$, that depends also on $T$ via
temperature dependence of parameter $a$ (see Eq. (14)) it appeared
cumbersome to calculate $T_{cr}$ and $R_{cr}$ analytically in general case.
Because of this we performed the calculation numerically introducing the
following dimensionless parameters and functions:

\begin{eqnarray}
R_{0} &=&\sqrt{\frac{\delta }{4\pi }},\text{\quad }\theta _{0}=\frac{\alpha
T_{c}}{4\pi },\text{\quad }\rho _{0}=\frac{\mu _{0}}{\pi R_{0}}\left( \frac{%
Q_{11}+2Q_{12}}{C_{11}+2C_{12}}\right) ,  \eqnum{20} \\
\lambda _{0} &=&\frac{\sqrt{4\pi \delta }}{\mu _{0}d^{2}},\text{\quad }\eta =%
\frac{\varkappa \,n\,q^{2}}{\mu _{0}},\quad \Delta \mu _{FE}(T,R)=\frac{%
\varepsilon _{i}-\varepsilon _{e}}{\varepsilon _{i}(\varepsilon
_{i}+\varepsilon _{e})}.  \nonumber
\end{eqnarray}
The results of $T_{cr}$ calculations is represented in Fig. 1.

The increase of $\mu _{0}$ value at other fixed parameters leads to the
decrease $\lambda _{0}$, that results in critical temperature decrease and
in critical radius increase (compare the curves 1 and 2 and their crossing
points with abscissa axis). Also one can see, that the decrease of
surrounding medium dielectric permittivity $\varepsilon _{e}$ leads to $%
R_{cr}$ decrease and $T_{cr}$ increase, these effects being larger for
smaller $\lambda _{0\text{ }}$value. Dashed-dotted curves correspond to
approximate analytical formulas obtained by the following way.

Estimations have shown, that for $\varepsilon _e\leq 100$ and $R\geq 10$ nm
the contribution of $a_0$ into $a$ value is the largest one, i.e. further we
can put $a\approx a_0$. Keeping in mind that $\sqrt{\delta /a}$ can be
estimated as correlation radius value, that has to decrease due to
depolarization field contribution, it is reasonable to suppose that $\sqrt{%
\delta /a_0}$ value is about several lattice constants. Therefore for $R%
\sqrt{a_0/\delta }>3\div 5$ one can derive from (19c), allowing for that at
transition point $\mu \rightarrow \mu _0$, the following approximate
analytical expression for the critical temperature of size driven phase
transition:

\begin{equation}
T_{cr}(R)\approx T_{c}\left( 1-\frac{R_{L}}{R}-\frac{R_{\lambda }}{R-R_{c}}%
\right)  \eqnum{21a}
\end{equation}
Here we introduce the parameters

\[
R_{c}=\frac{R_{0}\lambda _{0}\sqrt{4\pi }}{\lambda _{0}\sqrt{a_{0}}+\sqrt{%
4\pi }},\text{\quad }R_{L}=\rho _{0}\frac{R_{0}}{\theta _{0}},\quad
R_{\lambda }=\frac{R_{0}}{\theta _{0}}\left( \frac{3\sqrt{a_{0}}}{\lambda
_{0}\sqrt{a_{0}}+\sqrt{4\pi }}\right) . 
\]
$R_{c}$ can be considered as correlation radius renormalized by contribution
of depolarization field and surface tension. $R_{L}$ is correlation radius
renormalized only by surface tension.

For arbitrary $T<T_{c}$ approximate formula for critical radius $R_{cr}(T)$
can be obtained from Eq. (21a), as the solution of the quadratic equation
with respect to $T_{cr}=T,\_R=R_{cr}.$ The analytical expression can be
obtained for the some cases.

In particular, for the large enough particles when $R\sqrt{a_{0}/\delta }>>1$
and $R/R_{c}>>1,$ one can neglect $R_{c}$ in the denominator of the third
term of Eq. (21a) and formula for $T_{cr}(R),\_R_{cr}(T)$ dependences can be
written as

\begin{eqnarray}
T_{cr}(R) &\approx &T_{c}\left( 1-\frac{R_{L}+R_{\lambda }}{R}\right)
=T_{c}\left( 1-\frac{R_{cr}(T)}{R}\left( 1-T/T_{c}\right) \right) , 
\eqnum{21b} \\
R_{cr}(T) &\approx &\frac{R_{0}}{\theta _{0}(1-T/T_{c})}\left( \rho _{0}+%
\frac{3\sqrt{a_{0}}}{\lambda _{0}\sqrt{a_{0}}+\sqrt{4\pi }}\right) . 
\nonumber
\end{eqnarray}
If $R_{L}<<R_{\lambda }$ and $R_{\lambda }/\left( 1-T/T_{c}\right) <<R_{c}$
approximate formula for $T_{cr}(R),$ $R_{cr}(T)$ dependences can be written
as:

\begin{equation}
T_{cr}(R)\approx T_{c}\left( 1-\frac{R_{\lambda }}{R-R_{cr}}\right) ,\text{%
\quad }R_{cr}\approx R_{c}  \eqnum{21c}
\end{equation}
Thus, in this particular temperature range critical radius does not depend
on temperature.

It is seen from (21b) that $(T_{cr}-T_{c})$ is inversely proportional to the
particle size and e.g. for ferroelectrics with perofskite structure $%
T_{cr}<T_{c}$ because $Q_{11}+2Q_{12}>0$ (although $Q_{12}<0$). It should be
noted that the value of $(T_{cr}-T_{c})$ is proportional to the coefficient
of surface tension, the first and the second term in the brackets being
related to surface tension and surface energy respectively. One can see,
that Eq. (21b) describes the main features of $T_{cr}$ behavior: $%
T_{cr}=T_{c}$ at $\mu _{0}=0$, $T_{cr}$ decreases with $\mu _{0}$ increases
and $T_{cr}$ value increases with the particle size increases ($%
T_{cr}\rightarrow T_{c}$ at $R\rightarrow \infty $). The results of more
detailed calculation of $T_{cr}(R)$ dependence given by Eq. (21b) is shown
in Fig.1 by dashed-dotted line. One can see that the coincidence between
calculations on the base of Eq. (19c) and on the base of approximate formula
(21b) is very good for $\lambda _{0}=250,$ for $\lambda _{0}=25$ the $%
T_{cr}(R)$ dependences given by curve 1 and dash-dotted line are similar,
however quantitative values of $T_{cr}$ for both curves are different.

The equation (21b) at $T=0$ does correspond to the crossing points of the
curves in Fig. 1 with abscissa axis. But it is possible to obtain $R_{cr}$
at any fixed arbitrary temperature with the help of Eq. (21b).

It follows from Eqs. (21) that critical parameters of size driven
ferroelectric phase transition in nanomaterials are defined completely by
physical characteristics of the material such as Curie-Weiss constant $%
C_{W}=4\pi /\alpha $, electrostriction constants $Q_{ij}$, elastic modulus $%
C_{ij}$, piezoelectric constant $d_{1z}\equiv d$ and surface tension
coefficient $\mu _{0}$. This appeared possible due to the fact that
extrapolation length that usually is a phenomenological parameter was
expressed via $\mu _{0}$ and $d$ coefficients and surface tension was
related to polarization via electrostriction effect (see section 2).

\section{Properties in ferroelectric phase and free energy with renormalized
coefficients}

In ferroelectric phase $R>R_{cr}$, $T<T_{cr}$ the nonlinear term in Eqs.
(13), (15) can not be neglected. The simplest way to take it into account is
to look for polarization on the base of direct variational method. We choose
solution (19a)\ as a trial function and amplitude factor will be treated as
variational parameter, therefore

\begin{equation}
P^{FE}=P_V\left( 1-\frac Rr\frac{sh(r\sqrt{a/\delta })}{M(R)}\right) 
\eqnum{22}
\end{equation}
Here $P_V$ is variational parameter that represents the amplitude of
polarization space distribution.

After substitution of the trial function (22) into integral (13) one can
obtain the free energy in the following form:

\begin{equation}
F=\frac{A_{R}}{2}P_{V}^{2}+\frac{B_{R}}{4}P_{V}^{4}+\frac{C_{R}}{6}%
P_{V}^{6}-E_{R}P_{V}  \eqnum{23a}
\end{equation}
Here

\begin{eqnarray*}
A_{R} &=&\frac{3}{R^{3}}\int\limits_{0}^{R}r^{2}dr\left[ a\left( 1-\frac{R}{r%
}\frac{sh(r\sqrt{a/\delta })}{M(R)}\right) ^{2}+\delta \left( \frac{d}{dr}%
\left( \frac{R}{r}\frac{sh(r\sqrt{a/\delta })}{M(R)}\right) \right)
^{2}\right] - \\
&&-a_{0}\left( W(R)\right) ^{2}+\frac{\delta }{\lambda }\frac{3}{R}\left( 1-%
\frac{sh(R\sqrt{a/\delta })}{M(R)}\right) ^{2}
\end{eqnarray*}
\begin{eqnarray}
B_{R} &=&b\frac{3}{R^{3}}\int\limits_{0}^{R}r^{2}dr\left( 1-\frac{R}{r}\frac{%
sh(r\sqrt{a/\delta })}{M(R)}\right) ^{4},\qquad C_{R}=c\frac{3}{R^{3}}%
\int\limits_{0}^{R}r^{2}dr\left( 1-\frac{R}{r}\frac{sh(r\sqrt{a/\delta })}{%
M(R)}\right) ^{6},  \nonumber \\
E_{R} &=&E_{0}W(R),  \eqnum{23b}
\end{eqnarray}
where parameters $a$, $b$ and $c$ are given by Eq. (14).

Keeping in mind, that average polarization is $\overline{P}=P_{V}W(R)$, we
can rewrite Eq. (23a) as follows 
\begin{equation}
F=\frac{A_{R}}{W^{2}(R)}\frac{\overline{P}^{2}}{2}+\frac{B_{R}}{W^{4}(R)}%
\frac{\overline{P}^{4}}{4}+\frac{C_{R}}{W^{6}(R)}\frac{\overline{P}^{6}}{6}-%
\overline{P}E_{0}  \eqnum{23c}
\end{equation}

It is worth to underline that the terms in Eq. (13) which correspond to
polarization gradient, depolarization field and surface energy contribute to 
$A_R$ (see respectively the second, the third and the forth terms in $A_R$).
Therefore the main peculiarities of nanomaterials properties is expected to
be related to the first term in Eqs. (23a), (23c).

It is obvious that Eqs. (23a) and (23c)\ have the from of power series of
the amplitude of the polarization space distribution and average
polarization respectively. In general case all the coefficients (23b)\
depend on a particle size and temperature.

The main advantage of the Eqs. (23a,c), that they make it possible to obtain 
$P_V$ and $\overline{P}$ by conventional minimization procedure. In
particular for the phase transitions of the second order

\begin{equation}
P_V=\sqrt{-\frac{A_R}{B_R}},\quad \overline{P}=W(R)\sqrt{-\frac{A_R}{B_R}} 
\eqnum{24}
\end{equation}
and the polarization profile can be easily obtained with the help of Eq.
(22).

The physical properties e.g. dielectric susceptibility $\chi (r)$, $%
\overline{\chi }$ and pyrocoefficient $\Pi (r)$, $\overline{\Pi }$ can be
calculated as the derivative of polarization over $E_0$ and $T$ respectively.

The results of calculations of average dielectric susceptibility dependence
on the particle size and temperature are depicted in Figs. 2 and 3
respectively both for paraelectric ($R<R_{cr}$, $T>T_{cr}$) and
ferroelectric ($R>R_{cr}$, $T<T_{cr}$) phases for different values of
dimensionless extrapolation length $\lambda _{0}$. It is seen that inverse
susceptibility linearly depends on $T$ like that in Curie-Weiss (C-W) law
(see Fig. 3), although in ferroelectric phase there is small deviation from
linearity for small extrapolation length (see curves 1). In paraelectric
phase $1/\overline{\chi }\sim (R_{cr}/R-1)$ while in ferroelectric phase the
dependence of $\overline{\chi }$ on $R^{-1}$ declines from linearity more
strongly, especially for $\mu (T)/\mu _{0}-1=\eta \Delta \mu _{FE}(T)\neq 0$
(see solid curve 2), although general view resembles that of C-W law (see
inset to Fig. 2) with maximum at $R=R_{cr}$. It follows from Figs. 2,3 that
at $\mu _{0}=0$ $\overline{\chi }=\chi _{b}$.(because at $\mu
_{0}\rightarrow 0$ $\lambda _{0}\rightarrow \infty $, that corresponds to
the bulk) and the decrease of $\eta $ value makes the dependences linear,
i.e. $1/\overline{\chi }\sim (R_{cr}/R-1)$ and $1/\overline{\chi }\sim
(T-T_{c})$ (see dashed curves in the Figs.). Thus all deviation from
linearity in ferroelectric phase is related to $\Delta \mu _{FE}(T)$
contribution. The influence of dielectric permittivity $\varepsilon _{e}$ of
the surrounding medium appeared to be strong enough (compare solid and
dotted curves in Figs. 2,3). This effect is related to the contribution of
depolarization field and temperature dependent part of surface tension.

Practically the same type of influence of $\mu _{0}$, $\Delta \mu $ and $%
\varepsilon _{e}$ on the averaged spontaneous polarization one can see from
Figs. 4 and 5, where the temperature and size dependence of averaged
polarization are depicted, the crossing points of the curves with abscissa
axis in Figs. 4 and 5 being respectively $R_{0}/R_{cr}$ and $T_{cr}/T_{c}$
values. One can see, that the dependence of this quantities on extrapolation
length is a good agreement with approximate formulas (21b,c), where the
second term in the brackets reflects the influence of extrapolation length
i.e. the larger $\lambda _{0}$ the larger $T_{cr}$ and the smaller $R_{cr}$
value has to be. This behavior is related to the fact, that larger
extrapolation length corresponds to more homogeneous polarization like that
in bulk material, so that the closeness of $T_{cr}$ to $T_{c}$ has to be
expected. It is evident, that the destruction of such ''strong''
polarization can be achieved for small enough particles, i.e. for small $%
R_{cr}$ value. It is worth to underline, that the calculation of $T_{cr}$
and $R_{cr}$ with the help of Eqs. (21b,c) leads to the following values:\ $%
T_{cr}=0.44T_{c},$ $R_{cr}=0.036R_{0\text{ }}($for $\lambda _{0}=25)$, $%
T_{cr}=0.755T_{c},$ $R_{cr}=0.0785R_{0\text{ }}($for $\lambda =250)$, which
are cloth to those obtained from crossing points in Figs. 4, 5. This speaks
in favor of statement that the accuracy of formulas (21b,c) is not bad at
all.

The calculations of space distribution of polarization (see Eqs. (22), (24))
had shown, that the contribution of temperature dependent part of surface
tension can be large enough (compare solid and dashed lines in Fig. 6). It
is seen, that $\eta \Delta \mu _{FE}(T)\neq 0$ flattens the curves and
decreases polarization, these two effects can be related to increase of
extrapolation length and surface tension respectively with $\mu (T)/\mu
_0=1+\eta \Delta \mu _{FE}$ increase. Therefore the influence of surface
tension and its change under externally controllable parameters (e.g.
temperature) can manifest itself mainly in space distribution of physical
quantities, the smaller the particle or extrapolation length the larger the
effect can be (see Fig. 6).

The considered temperature and size dependences of properties were
calculated numerically because of the complex form of the coefficients
(23b). These dependences can be more clearly seen after analytical
calculation of integrals (23b). In particular for $A_R$ the integration
yields :

\begin{equation}
A_R=\left[ a-a_0W(R)\right] W(R).  \eqnum{25}
\end{equation}
It is seen that expression in the brackets completely coincides with left
hand side of Eq. (19c), so that at $T=T_{cr}$ or $R=R_{cr}$ $A_R=0$, $P_V=0$%
, $\overline{P}=0$ as it has to be.

The expressions for $B_{R}$ and $C_{R}$ appeared to be very complex (see
Appendix 2). But all the coefficients can be simplified allowing for the
estimations made in previous section for $T_{cr}$ and $R_{cr}$ approximate
calculations. Since the accuracy of these formulas was shown to be good, it
is worth to write the approximate formulas for the coefficients $A_{R}$, $%
B_{R}$ and then for the physical properties. Allowing for the approximate
expression for $W(R)$ at $R\sqrt{a_{0}/\delta }>3\div 5$ one can rewrite $%
A_{R}$ approximately in the following form: 
\begin{equation}
A_{R}\approx \alpha (T-T_{c})+\frac{\mu }{R}\left[ 4\frac{Q_{11}+2Q_{12}}{%
C_{11}+2C_{12}}+3\frac{d^{2}\sqrt{\delta a_{0}}}{d^{2}\mu _{0}+\sqrt{\delta
a_{0}}}\right] .  \eqnum{26a}
\end{equation}
Substitution of Eqs. (21b) for $T_{cr}$ or $R_{cr}$ into Eq. (26a) yields 
\begin{eqnarray}
A_{R} &=&\alpha \left( T-T_{cr}+\frac{\Delta \mu _{s}}{\mu _{0}}%
(T_{c}-T_{cr})\right) ,  \eqnum{26b} \\
A_{R} &=&\alpha (T-T_{c})\left( 1-\frac{\mu _{0}+\Delta \mu _{s}}{\mu _{0}}%
\frac{R_{cr}}{R}\right) .  \eqnum{26c}
\end{eqnarray}

One can see, that for the case when $\Delta \mu _{s}/\mu _{0}\ll 1$
expression (26b) corresponds to that for C-W law, and (26c) contains the
simple multiplier $(1-R_{cr}/R)$. But in general case situation is more
complicated. Indeed, because $\Delta \mu _{s}$ depends on the material
dielectric permittivity (see Eq. (6a)) it can be expressed via $A_{R}$. In
particular it is possible when $(\varepsilon _{i}-\varepsilon
_{e})/(\varepsilon _{i}(\varepsilon _{i}+\varepsilon _{e}))\approx
1/\varepsilon _{i}$, that is proportional to $A_{R}$. In such a case (26b)
and (26c) can be represented as 
\begin{equation}
A_{R}=\alpha (T-T_{cr})A_{R1}  \eqnum{27a}
\end{equation}
\begin{equation}
A_{R}=\alpha (T-T_{c})\left( 1-\frac{R_{cr}}{R}\right) A_{R2}  \eqnum{27b}
\end{equation}
\begin{equation}
A_{R1}=\frac{1}{1-\alpha \eta (T_{c}-T_{cr})},\quad A_{R2}=\frac{1}{1+\alpha
(T-T_{c})\eta R_{cr}/R}  \eqnum{27c}
\end{equation}
Allowing for that e.g. for BaTiO$_{3}$ $\eta \approx 10^{2}$ and $\alpha
\sim 10^{-5}$ one can see, that $A_{R1}\approx 1$ and $A_{R2}\approx 1$
although the accuracy of this estimation for $A_{R1}$ is better than for $%
A_{R2}$. In what follows we will use Eqs. (27a), (27b) with $A_{R1}=A_{R2}=1$%
.

The formulas (27a) and (27b) can be applied respectively to study the
temperature dependence for some fixed radius of the particles and to study
the size dependence at some arbitrary fixed temperature.

Keeping in mind, that in adopted approximation $a\approx a_0$ and so it is
independent on temperature, the coefficient $B_R$ also independent on $T$
(see Eq. (23b) and Appendix 2), so that the properties temperature
dependence defines completely by (27a). Moreover, because $W(R)\approx 1$,
so $P_V\approx \overline{P}$ (see Eq. (24)) and $E_R\approx E_0$ (see Eq.
(23b)). In such a case the susceptibility has the conventional form

\begin{equation}
\overline{\chi }(T)=\frac 1{\alpha (T-T_{cr})},\quad T>T_{cr}  \eqnum{28a}
\end{equation}
and

\begin{equation}
\overline{\chi }(T)=\frac{1}{2\alpha (T_{cr}-T)},\quad T<T_{cr}  \eqnum{28b}
\end{equation}
respectively for paraelectric and ferroelectric phase of size driven
transition. The temperature dependences given by Eqs. (27a), (28a), (28b)
correspond completely to those in bulk materials but with substitution $%
T_{cr}$ for $T_{c}$. These dependences are shown in the Fig. 3 by
dash-dotted curves. One can see, that the approximate formulas describe the
accurate curves good enough, especially curves numbered by 2, where
dash-dotted curve completely coincides with solid curve, although the
coincidence is not so good for the curves 1. Taking into account, that for
ferroelectrics $\varepsilon _{i}\approx 4\pi \chi ,$ it is easy to obtain
after substitution of $T_{cr}(R,R_{cr})$ from Eqs.\_(21b) into Eq.\_(28),
that at fixed temperature $T$ the dependence of nanoparticle dielectric
permittivity over particle radius $R$ can be represented as:

\begin{eqnarray}
\overline{\varepsilon _i}(T,R) &=&\frac{\varepsilon _b}{2\left(
1-R_{cr}(T)/R\right) },\quad R>R_{cr}(T),  \eqnum{28c} \\
\overline{\varepsilon _i}(T,R) &=&\frac{\varepsilon _b}{\left(
R_{cr}(T)/R-1\right) },\quad R<R_{cr}(T).  \nonumber
\end{eqnarray}

Here $\varepsilon _b=C_W/(T_c-T).$

The spontaneous polarization approximate formulas can be obtained from Eq.
(24) at $W(R)\approx 1$ and $B_R\approx \beta $ (see Appendix 2). This
yields for average polarization

\begin{equation}
\overline{P}=\sqrt{\frac \alpha \beta (T_{cr}-T)}  \eqnum{29}
\end{equation}

This formula completely coincides with that for bulk material when
substituting $T_{cr}$ for $T_c$. The approximate expression (29) describes
exact curves $\overline{P}(T)$ not bad at all (compare solid and dash-dotted
curves in Fig. 5 which completely coincide for $\lambda _0=250$). Note, that
these curves describe also the temperature dependence of inverse
pyrocoefficient $\overline{\Pi }$ because

\begin{equation}
\overline{\Pi }=\frac{d\overline{P}}{dT}=-\frac \alpha \beta \frac 1{%
\overline{P}}  \eqnum{30}
\end{equation}

For the phase transitions of the first order the spontaneous polarization,
the dielectric susceptibility and the pyrocoefficient depend also on $C_R$
coefficient given by (23b). It was shown (see Appendix 2) that $C_R\approx c$%
, so that all the physical quantities for nanomaterials can be described by
the same formulas as for bulk materials, but with the coefficient before $%
P^2 $ in the form of (27a) or (27b). The substitution of Eq. (27b) will make
it possible to describe approximately size dependence of the properties. The
description on the base of the approximate formulas appeared to be good
enough especially for the curves 2 (compare solid and dash-dotted curves in
Figs. 2, 4).

The space distribution of the properties in nanomaterial can be calculated
with the help of Eq. (22). Since in adopted approximation $a\approx a_0$ is
temperature independent quantity and $\overline{P}=P_V$ the properties
profiles will be described by the expression in the brackets of Eq. (22),
i.e. their form will be the same as that depicted in Fig. 6.

\section{The depolarization field and surface tension coefficient size and
temperature dependencies}

In this section we will consider in details size and temperature effects in
depolarization field and surface tension coefficient.

\subsection{The depolarization field}

It was shown in Sect. 3, that for nanoparticles with spherical form the
depolarization field is given by Eq. (12) in the model, when it can be
partly compensated by space-charge layer. If dielectric permittivity of
surrounding media $\varepsilon _{e}$ independent on $T$, temperature
dependence of $E_{dz}\equiv \dot{E}_{d}$ will be related to that of $(-P_{z}+%
\overline{P}_{z})\equiv (-P+\overline{P})$. Allowing for space distribution $%
P(r)$ (see Eq. (22)) one can calculate $E_{d}(r)$ distribution, which is
depicted in the inset to Fig. 6. It is seen that for infinitely large
extrapolation length ($\lambda _{0}\sim 1/\mu _{0},~\mu _{0}=0$) that
corresponds to practically homogeneous polarization, $E_{d}=0$. For smaller
extrapolation length values when polarization is inhomogeneous $E_{d}<0$
inside the particle and transforms into $E_{d}>0$ near the surface. The
absolute values of $E_{d}$ decrease with $\lambda _{0}$ increase (compare
the curves 1 and 2 in the inset to Fig.6). Therefore depolarization field
tends to decrease the polarization inside the particle and to increase it
near the surface. Since in the considered case of positive extrapolation
length polarization near the surface is smaller than inside the particle,
depolarization field tends to flatten the polarization in nanoparticle, i.e.
to make it more homogeneous.

\subsection{Surface tension coefficient}

Let us discuss now the size and temperature dependence of surface tension
coefficient in ferroelectric nanomaterials. In the adopted model surface
tension coefficient is given by Eqs. (7), (6), so that its size and
temperature dependences are defined by that of nanoparticle dielectric
permittivity $\varepsilon _i$ in supposition that permittivity of the
surrounding media $\varepsilon _e$ is temperature independent quantity.
Because of $\varepsilon _i$ dependence on the temperature $T$ and particle
radius $R$ the quantities $\Delta \mu _s$ and $\Delta \mu _{FE}$ from Eqs.
(6), (20c) have to depend on these parameters. One can see the dependence $%
\Delta \mu _{FE}(T)$ in Fig. 7, and $\Delta \mu _{FE}(R)$ in Fig.8 for the
several values of $\varepsilon _e$. Negative sign of $\Delta \mu _{FE}$
corresponds to the case $\varepsilon _i<\varepsilon _e$, that realizes at $%
\varepsilon _e\geq 300$ both for $\lambda _0=25$ and $250$ (see inset). The
existence of the maximum in curve 3 is related to the temperature region
where $\varepsilon _i$ and $\varepsilon _e$ is close to one another.
Positive sign of $\Delta \mu _{FE}$ corresponds to the case $\varepsilon
_i>\varepsilon _e$ (see curves 1, 2), the curve 1 being very close to the
respective curve for $1/\overline{\chi }$ dependences on temperature (see
Fig.3) and on the particle radius (see inset to Fig. 8).

One can see from Figs.7, 8 that $\left| \Delta \mu _{FE}\right| \sim
10^{-3}\div 10^{-2}$, i.e. the relative contribution of $\Delta \mu _{s}$ to
the surface tension $\mu $ is equals $\Delta \mu _{s}/\mu _{0}=\eta \Delta
\mu _{FE}\symbol{126}10^{-1}\div 1,\ $for $\eta =100$ that we used when
plotting all the Figures 1-6.

Therefore for nanomaterials $\Delta \mu _s(T)$ can be both positive and
negative quantity, its absolute value $\left| \Delta \mu _s(T)\right| \leq
\mu _0$. This different possibilities depend on the material characteristics
(particles size, extrapolation length, transition temperature in bulk) and
dielectric constant of the surrounding media.

The temperature dependence of surface tension coefficient results in
extrapolation radius temperature dependence because $\lambda \sim 1/\mu $
(see Eq. (15c)). Since in paraelectric phase $\mu =\mu _{0}$ is temperature
independent quantity, we depicted $\lambda /\lambda _{PE}=1/(1+\Delta \mu
_{s}/\mu _{0})$ in Fig. 9 as the function of temperature. Note, that $%
\lambda _{PE}=R_{0}\lambda _{0}.$ One can see, that in paraelectric phase ($%
T\geq T_{cr}$) $\lambda /\lambda _{PE}=1$, while in ferroelectric phase ($%
T<T_{cr}$) its behavior depends strongly on $\varepsilon _{e}$ value as it
has been already discussed above, allowing for $\Delta \mu _{s}/\mu
_{0}=\eta \Delta \mu _{FE}$. The same type behavior was obtained for $%
\lambda /\lambda _{PE}$ dependence on particle size with $\lambda /\lambda
_{PE}=1$ at $R\leq R_{cr}$ (see inset 1 to Fig. 9). Note, that in the case $%
\Delta \mu _{s}>0$ the extrapolation length value decreases. Under the
condition when $\Delta \mu _{s}<0$ extrapolation length has to increase
resulting into more homogeneous polarization in the nanoparticle.

\section{Discussion}

1. Let us compare our theoretical calculations with experimental dependences 
$T_{cr}(R)$ and $\overline{\varepsilon _i}(R)$ for BaTiO$_3$ and PbTiO$_3$
nanomaterials \cite{Uchino}, \cite{Ishikawa}, \cite{Rag}$.$

Taking into account that $R_\lambda \sim 1/\sqrt{1+2\varepsilon _e}$ (see
Eqs. (12), (20),(21a)), i.e. decreases when $\varepsilon _e$ increases, and $%
R_c$ saturates when $\varepsilon _e$ increases, one can obtain that two
cases are possible. For ceramics with $\varepsilon _e>>1,$ one can suppose
that $R_{\lambda ,c}<<R_L$ and use Eqs. (21b), (28c). In contrast, for
powder samples $\varepsilon _e\sim 1,$ one can suppose that $R_\lambda >>R_L$
and therefore Eqs. (21c) is valid. Really, the estimation of the parameters
for aforementioned materials (see, e.g. \cite{JonaShirane}, \cite{PeTa}, 
\cite{QuZhPr}) have shown that Eqs.(21b) and (21c) can be used for the
ceramic and powders description respectively.

Below we regard critical radius as the fitting parameter, so Eqs. (28c)
depend on the one fitting parameter $R_{cr}=R_{cr}(T)$ at fixed room
temperature $T$. The comparison of calculated on the base of Eqs. (28c) with
the experimental $\overline{\varepsilon _i}$($R$) dependence for BaTiO$_3$
ceramics is represented in Fig. 10. It is seen good agreement between
experimental and calculated data. The difference between obtained critical
radii value can be the result of difference in intergrain media due to
different technological processes of BaTiO$_3$ ceramics production (see \cite
{Niep}, \cite{Rag} and ref. therein) which can result in difference of
surface tension coefficients.

Critical temperature $T_{cr}(R)$ depends on the two parameters: radii $R_{cr}
$ and $R_\lambda $. The comparison of (21c) with the experimental dependence 
$T_{cr}(R)$ for powder BaTiO$_3$ and PbTiO$_3$ samples is represented in
Fig. 11. One can see that the theory fits the experimental data rather well.

Using the fitting parameters values, given in the captions to figures 10 and
11, the known values of the electrostriction, elastic stifness, Curie-Weiss
constants and critical temperatures (see, e.g. \cite{JonaShirane}, \cite
{PeTa}) and the following values of the polarization gradient coefficients
for BaTiO$_3$ and PbTiO$_3$ in CGSE-units (see, e.g. \cite{QuZhPr}): $\delta
($BaTiO$_3)=5\,\cdot 10^{-15}cm^2,\quad \delta ($PbTiO$_3)=5\,\cdot
10^{-16}cm^2,$one can obtain from Eqs. (15c),(21a), the the following data
for physical quantities used in the theory: $\mu _0=3.11\ \cdot
10^5\,din/cm^2,\quad \lambda _{PE}=16\,nm$ (BaTiO$_3$ ceramics with $%
\varepsilon _e=100$ and $d=10^{-7}$ CGSE-units); $\mu _0=3.28\ \cdot
10^5\,din/cm^2,\quad \lambda _{PE}=50\,nm$ (BaTiO$_3$ powder with $%
\varepsilon _e=1$ and $d=0.55\,\cdot 10^{-7}$ CGSE-units). Using the known
value of the surface tension of PbTiO$_3$ powder $\mu _0=5\ \cdot
10^4\,din/cm^2$ \cite{Wenhui} and $\varepsilon _e=1$ one can obtain $d=1.20\
\cdot 10^{-7}\ $CGSE\ units, $\lambda _{PE}=6.9\,nm.$

One can see that the obtained quantities have reasonable values known for
these parameters (see, e.g. \cite{Uchino}, \cite{Ishikawa}, \cite{QuZhPr}).

2. The surface tension and depolarization field influence were taken into
account in the thermodynamic approach to the investigation of ferroelectric
nanomaterials properties. Surface energy was expressed via surface tension
coefficient that results into extrapolation length dependence on this
coefficient. Its value was obtained in the framework of the model taking
into account the space-charge layer on the particle surface and shown to be
temperature dependent quantity. In the same model the depolarization field
was taking into consideration. This field was shown to be partially
compensated by the surface charges and proportional to the deviation of the
polarization from its mean value. Coefficient of proportionality is defined
by the nanoparticle shape and surroundings dielectric permittivity. This
lead to the essential difference between dielectric properties of the
ceramic and powder samples because of the surrounding media properties
difference.

The study of the physical properties of the ferroelectric nanomaterials such
as transition temperature, critical particle size, spontaneous polarization
and dielectric susceptibility were performed on the base of Euler-Lagrange
equation for the polarization with the boundary condition with the
temperature dependent extrapolation length of the positive sign. It should
be underlined that its negative sign can arise from the negative sign of the
coeffient before square polrization gradient. But in this case the higher
power gradient terms has to be included into the free energy in order to
stabilize the system.

The possibility to calculate physical properties of the nanomaterials by the
minimization of the ''bulk'' free energy density with renormilized
coefficients, which depend on the particle size, temperature, surface
tension and depolarization field characteristics has been demonstrated. The
developed theory gives the basis to the empirical expression proposed in 
\cite{Uchino} for the BaTiO$_3$ and PbTiO$_3$ transition temperature
dependences on the particles size and fits rather well the available
experimental data.

\section{Appendix 1}

Let us consider the point charge $q$ located at the distance $r_0$ from the
center of the dielectric sphere with radius $R>r_0$ and dielectric
permittivity $\varepsilon _i$. The outer space is also regarded as
dielectric with permittivity $\varepsilon _e.$

In order to find the electric potential $\varphi _q^{i,\ e}$ of this system
one have to solve Laplace equation with the appropriate boundary conditions.
In this case it has the following form: 
\begin{equation}
\Delta \varphi _q^{i,\ e}=0,\quad \left. \varphi _q^i\right| _{r=R}=\left.
\varphi _q^e\right| _{r=R},\quad \left. \varepsilon _i\nabla \varphi
_q^i\right| _{r=R}=\left. \varepsilon _e\nabla \varphi _q^e\right| _{r=R} 
\eqnum{A1-1}
\end{equation}

Having solved this equation in the spherical coordinates it is easy to
derive the surface contribution to the potential $\varphi _{q}^{i}$: 
\begin{equation}
\varphi _{q}^{s}(r_{0},r,\theta )=q\frac{\varepsilon _{i}-\varepsilon _{e}}{%
\varepsilon _{i}}\sum\limits_{m=0}^{\infty }\frac{m+1}{m\ \varepsilon
_{i}+(m+1)\varepsilon _{e}}\frac{r_{0}^{m}r^{m}}{R^{2m+1}}P_{m}\left( \cos
\left( \theta \right) \right) ;\quad r\leq R  \eqnum{A1-2}
\end{equation}
where $(r,\theta ,\varphi )$ are spherical coordinates and $P_{m}\left( \cos
\left( \theta \right) \right) $ is Legendre polynomial of $m$-th order. This
potential is determined by the surface bound charges induced by charge $q$.
Thus the additional surface energy can be calculated as interaction energy
between charge q and induced charges. Considering the electric dipole with
the dipole moment $q\,r_{d}$ directed along polar axes, one can write the
interaction energy with respect to $P_{m}\left( \cos \left( \theta =0\right)
\right) =1$ as follows 
\begin{eqnarray}
w_{s}(r_{0}) &=&q\left( (\varphi _{q}^{s}(r_{0},r_{0},0)-\varphi
_{q}^{s}(r_{0}+r_{d},r_{0},0))-(\varphi
_{q}^{s}(r_{0},r_{0}+r_{d},0)-\varphi
_{q}^{s}(r_{0}+r_{d},r_{0}+r_{d},0))\right) \approx  \nonumber \\
\ &\approx &(q\,r_{d})^{2}\frac{\varepsilon _{i}-\varepsilon _{e}}{%
\varepsilon _{i}}\sum\limits_{m=0}^{\infty }\frac{m+1}{m\ \varepsilon
_{i}+(m+1)\varepsilon _{e}}\frac{m^{2}r_{0}^{2m-2}}{R^{2m+1}}  \eqnum{A1-3}
\end{eqnarray}
Here we used the strong inequality $r_{d}<<r_{0}$. Then the surface energy
can be represented in the form of the equations (5a) (5b) 
\[
\Delta \mu ^{s}=n\frac{1}{R^{2}}\int\limits_{0}^{R-a}w_{s}(r_{0})\,r_{0}^{2}%
\,dr_{0} 
\]
Having performed the integration, one can obtain 
\begin{equation}
\Delta \mu ^{s}=n\frac{(q\,r_{d})^{2}}{R^{2}}\frac{\varepsilon
_{i}-\varepsilon _{e}}{\varepsilon _{i}}\sum\limits_{m=0}^{\infty }\frac{m+1%
}{m\ \varepsilon _{i}+(m+1)\varepsilon _{e}}\frac{m^{2}}{2m+1}\left( 1-\frac{%
a}{R}\right) ^{2m+1}  \eqnum{A1-4}
\end{equation}
The majorant series of the series from (A1-4) is equal to 
\begin{equation}
\frac{1}{2(\varepsilon _{i}+\varepsilon _{e})}\sum\limits_{m=0}^{\infty
}\left( m+1\right) \,b^{m}\left( 1-\frac{a}{R}\right)  \eqnum{A1-5}
\end{equation}
Here $b=\left( 1-a/R\right) ^{2}$. Sum of the series (A1-5) can be find in
the following way 
\[
\sum\limits_{m=0}^{\infty }\left( m+1\right)
\,b^{m}=\sum\limits_{m=0}^{\infty }\frac{d(b^{m+1})}{d\,b}=\frac{d}{d\,b}%
\left( \sum\limits_{m=0}^{\infty }b^{m+1}\right) =\frac{d}{d\,b}\left( \frac{%
b}{1-b}\right) =\frac{1}{(1-b)^{2}} 
\]
Using this relation it is easy to find the evident dependence of the series
(A1-5) on $a/R$ ratio. Owing to the inequality $a<<R$ we finally get the
following expression for the additional surface tension (A1-4)

\begin{equation}
\Delta \mu ^s=n\frac 18\left( \frac{r_d}aq\right) ^2\frac{\varepsilon
_i-\varepsilon _e}{\varepsilon _i(\varepsilon _i+\varepsilon _e)}. 
\eqnum{A1-6}
\end{equation}

\section{Appendix 2}

Let us proceed to the integration in the equations (23b).Owing to the
boundary condition (15b) integration by parts of the second term in $A_{R}$
leads to the following equation for the coefficient $A_{R}$%
\begin{eqnarray}
A_{R}+a_{0}\left( W(R)\right) ^{2} &=&\frac{3}{R^{3}}\int%
\limits_{0}^{R}r^{2}dr\left( 1-\frac{\sinh (r\sqrt{a/\delta })}{M(R)\ r/R}%
\right) \left[ a\left( 1-\frac{\sinh (r\sqrt{a/\delta })}{M(R)\ r/R}\right)
+\right.  \nonumber \\
\left. +\delta \left( \frac{R}{r}\frac{d^{2}}{dr^{2}}\frac{\sinh (r\sqrt{%
a/\delta })}{M(R)}\right) \right] &=&a\frac{3}{R^{3}}\int%
\limits_{0}^{R}r^{2}dr\left( 1-\frac{\sinh (r\sqrt{a/\delta })}{M(R)\ r/R}%
\right)  \eqnum{A2-1}
\end{eqnarray}
Since this integral is equal to $W(R)$ one can get the expression (25) for
the coefficient $A_{R}$. For the sufficiently large particles radius $R>>%
\sqrt{\delta /a}$ it is easy to substitute $\cosh (x)$ and $\sinh (x)$ for $%
\exp (x)/2$ and to rewrite expression for $W(R)$ as follows: 
\begin{equation}
W(R)=1-\frac{3}{\xi \left( \ell +1\right) -\ell }  \eqnum{A2-2}
\end{equation}
Here 
\[
\xi =R\sqrt{a/\delta }\text{,\quad }\ell =\lambda \sqrt{\frac{a}{\delta }}. 
\]

Integration in the expression for $B_{R}$ from the equation (23b) leads to
the following coefficient form 
\begin{eqnarray}
\frac{B_{R}}{b} &=&1-\frac{1}{8\xi ^{2}M(R)^{4}}\left( 9\xi ^{2}+72\xi
^{2}M(R)^{2}+96\varphi ^{3}\left( \xi \cosh (\xi )-\sinh (\xi )\right)
-\right.  \nonumber \\
&&-12\xi ^{2}\cosh (2\xi )+3\xi ^{2}\cosh (4\xi )-36\xi \,M(R)^{2}\sinh
(2\xi )-  \nonumber \\
&&\left. -72\xi ^{2}M(R)%
%TCIMACRO{\limfunc{Sih} }
%BeginExpansion
\mathop{\rm Sih}%
%EndExpansion
(\xi )+24\xi ^{3}%
%TCIMACRO{\limfunc{Sih} }
%BeginExpansion
\mathop{\rm Sih}%
%EndExpansion
(2\xi )+24\xi ^{2}M(R)%
%TCIMACRO{\limfunc{Sih} }
%BeginExpansion
\mathop{\rm Sih}%
%EndExpansion
(3\xi )-12\xi ^{3}%
%TCIMACRO{\limfunc{Sih} }
%BeginExpansion
\mathop{\rm Sih}%
%EndExpansion
(4\xi )\right)  \eqnum{A2-3}
\end{eqnarray}
Here and $%
%TCIMACRO{\limfunc{Sih}}
%BeginExpansion
\mathop{\rm Sih}%
%EndExpansion
(\xi )=\int\nolimits_{0}^{\xi }d\zeta \sinh (\zeta )/\zeta $ is the integral
hyperbolic sine function. When using the strong inequality (21c) in the form 
$\xi >>1$ and the asymptotic form of $%
%TCIMACRO{\limfunc{Sih}}
%BeginExpansion
\mathop{\rm Sih}%
%EndExpansion
(\xi )\approx \exp (\xi )/(2\,\xi )$ one can suppose that $M(R)=\exp (\xi
)(\ell .+1)/2$ and obtain the following expression for $B_{R}$ 
\begin{equation}
B_{R}\approx b\left( 1-\frac{3}{\xi \ (\ell +1)^{2}}\left( 4\ell +1\right)
\right)  \eqnum{A2-4}
\end{equation}
It is seen that under the condition 
\begin{equation}
\lambda \sqrt{\frac{a}{\delta }}=\ell >>1  \eqnum{A2-5}
\end{equation}
coefficient $B_{R}$ is close to $b$.

Coefficient $C_{R}$ also can be expressed in terms of the hyperbolic
functions and the integral hyperbolic sine function, but the final
expression is very cumbersome and omitted for the sake of briefness. Using
inequalities (21c) and (A2-5) we can suppose that

\[
1-\frac{R}{r}\frac{\sinh (r\sqrt{a/\delta })}{M(R)}\approx 1-\frac{R}{r}%
\frac{2\sinh (r\sqrt{a/\delta })}{\exp (\xi )\left( \ell +1\right) }\approx
1 
\]
Therefore when using the explicit form (23b) of the coefficients $B_{R}$ and 
$C_{R}$ their approximate expression can be easily simplified to

\begin{eqnarray*}
B_R &\approx &b \\
C_R &\approx &c
\end{eqnarray*}
The first of these approximate equalities coincides with (A2-4) owing to
strong inequality (A2-4).

\begin{center}
FIGURE CAPTIONS
\end{center}

{\bf Figure 1}. The size dependence of transition temperature 
at $\theta _0=0.005$, $\varepsilon _e=95$ and $\rho _0=0.05$
for different $\lambda _0$ and $\eta $: $\lambda _0=25$ (curves 1),$\lambda
_0=250$ (curves 2), $\eta =100$ (solid curves), $\eta =0$ (dashed curves), $%
\varepsilon _e=0$ (dotted curves), approximate formulas (dash-dotted
curves), $\mu _0=0$ (dash-double-dotted curves). Inset: the same dependences
over the inverse radius.

{\bf Figure 2}. The dependences of the inverse static susceptibility on inverse
radius ($\chi _b$ is bulk
static susceptibility) at fixed temperature $\theta _0(T/T_C-1)=-0.005$.
Inset: the same dependences of $\overline{\chi }/\chi _b$ on $R/R_0$%
. The parameters for the different curves correspond to those in Fig.1.

{\bf Figure 3}. The temperature dependences of inverse and direct (inset) 
static susceptibility at fixed radius $R/R_0=50$. 
The parameters for the different curves correspond to those in Fig.1.

{\bf Figure 4}. The size dependence of spontaneous polarization on
the inverse and direct (inset) particle radius ($P_{Sb}$ is bulk polarization) 
at $\theta _0(T/T_C-1)=-0.005$.  
The parameters for the different curves correspond to those in
Fig.1.

{\bf Figure 5}. The temperature dependences of spontaneous polarization 
over $T/T_c$ ($P_{Sb}^0$ is bulk polarization at zero temperature) at
fixed radius $R/R_0=50$ and $\theta _0(T/T_C-1)=-0.005$. The parameters for
the different curves correspond to those in Fig.1.

{\bf Figure 6}. The spatial distribution of the spontaneous polarization 
and the depolarization field (inset)  
at fixed temperature $\theta _0(T/T_C-1)=-0.005$, $R/R_0=40$. The
parameters for the different curves correspond to those in Fig.1.

{\bf Figure 7}. The temperature dependence of surface tension coefficient
at fixed radius $R/R_0=50$, $\theta _0=0.005$, $\lambda _0=25$ and $\rho
_0=0.05$, $\eta =100$ and different $\varepsilon _e$ values: $\varepsilon _e$%
=1 (curves 1), $\varepsilon _e$=100 (curves 2), $\varepsilon _e$=300 (curves
3), $\varepsilon _e\rightarrow \infty $ (curves 4). Inset: the same
dependence for $\lambda _0=250$.

{\bf Figure 8}. The size dependence of surface tension coefficient 
$\Delta \mu _{FE}$,
parameters are the same as in basic plot in Fig.7. Inset: the same
dependences for $1/\overline{\chi }$ in FE-phase.

{\bf Figure 9}. The temperature dependence of extrapolation length
(basic plot), parameters are the same as at basic plot in Fig.7. Inset: the
size dependence of extrapolation length at
fixed temperature $\theta _0(T/T_C-1)=-0.005$.

{\bf Figure 10}. The comparison of experimental (open squares \cite{Niep},
triangles \cite{Rag}) and theoretical given by Eq. (28c) (solid curves) 
dependences of the
static permittivity (basic plot) and impedance 
(inset) over particle diameter $D$ for BaTiO$_{3}$ ceramics. 
The fitting parameters: $\varepsilon_{B}=1450$, $D_{cr}=700$ nm (basic plot), 
$D_{cr}=1500$ nm (inset).

{\bf Figure 11}. The comparison of experimental (open squares \cite{Uchino}, 
\cite{Ishikawa}) and theoretical given by Eq. (21c) (solid curves) dependences 
of transition
temperature $T_{CR}$ over particle diameter $D$ for powder BaTiO$_{3}$ (basic
plot) and PbTiO$_{3}$ (inset) samples. The fitting parameters:
$D_{\lambda}=17.4$ nm, $D_{cr}=1000$ nm (BaTiO$_3$); $D_{\lambda}=8500$ nm,
$D_{cr}=12$ nm (PbTiO$_3$).

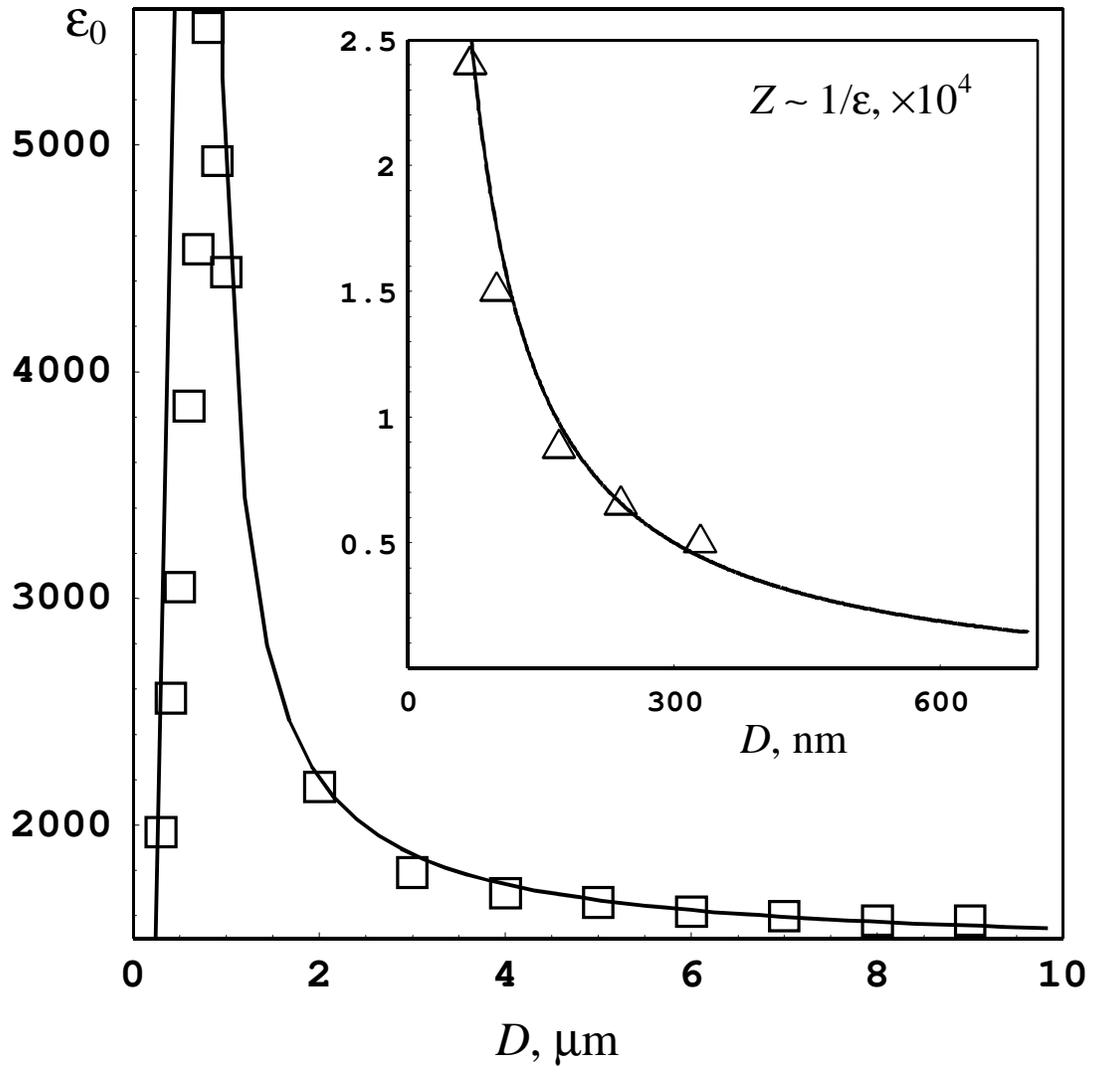
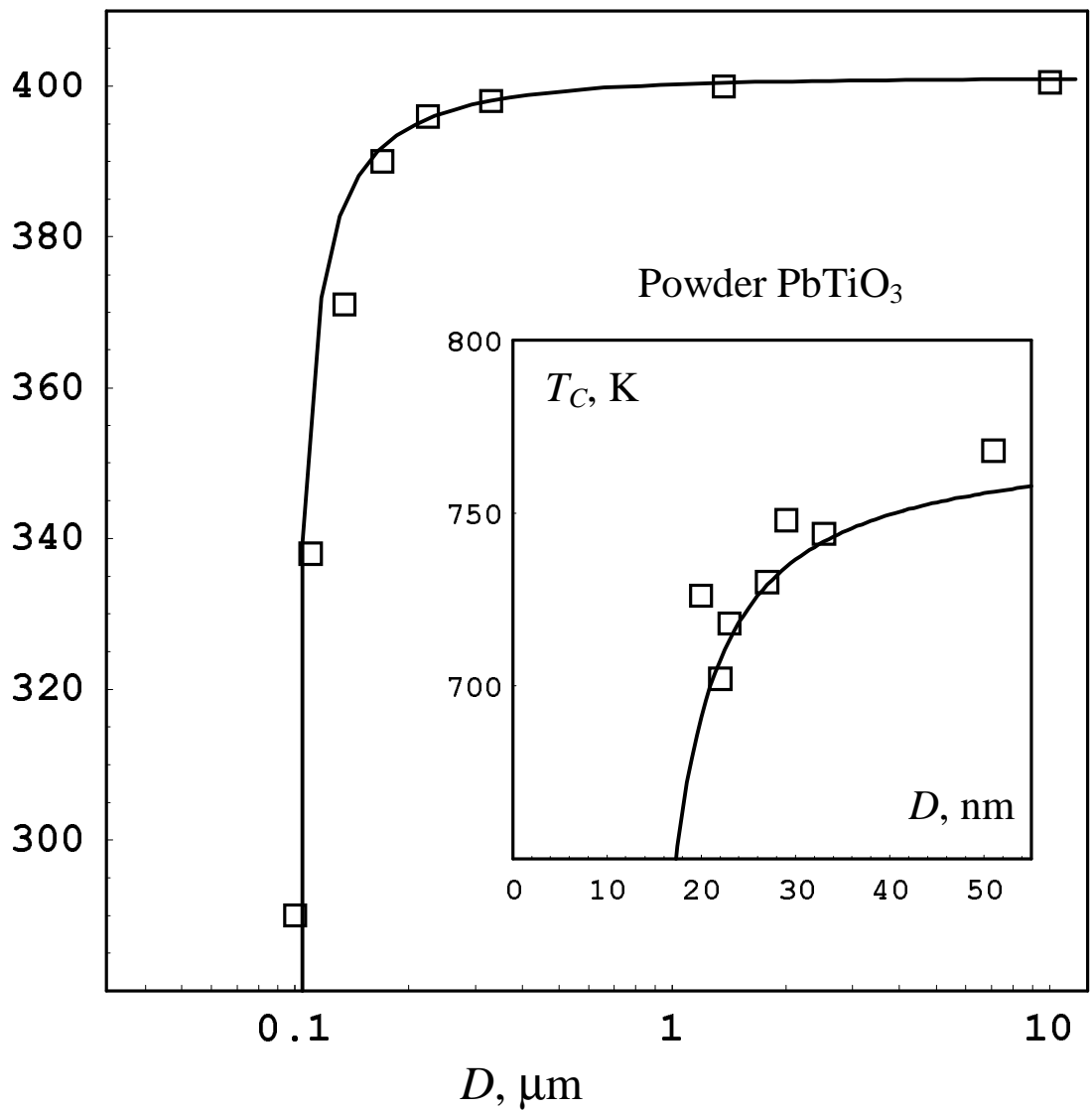
\end{document}